%%%%%%%%%%%%%%%%%%%%%%%  filename = ppt.lanl.ellis.tex  %%%%%%%%%%%%%%%%%%%%%%%
%                                                                             %
%  This source file is written in TeX and uses AMSTeX macros.                 %
%                                                                             %
%%%%%%%%%%%%%%%%%%%%%%%%%%%%%%%%%%%%%%%%%%%%%%%%%%%%%%%%%%%%%%%%%%%%%%%%%%%%%%%

\nopagenumbers
\input amstex

\def\degree{\mathaccent"17 {}}

\def\Ghat{\hat G}
\def\epshat{\hat \epsilon}
\def\bdhat{\, @, @, @, @, @,{\hat {\! @! @! @! @! @!{\bold d}}}}
\def\omegahat{\hat \omega}
\def\tauhat{\hat \tau}

\def\Ao{\mathaccent"17 A}
\def\Co{\mathaccent"17 C}
\def\Fo{\mathaccent"17 F}
\def\go{\mathaccent"17 g}
\def\Go{\mathaccent"17 G}
\def\mo{\mathaccent"17 m}
\def\Po{\mathaccent"17 P}

\def\tauo{\mathaccent"17 \tau}
\def\phio{\mathaccent"17 \phi}

\def\Gammao{\mathaccent"17 \Gamma}
\def\Deltao{\mathaccent"17 \Delta}

\def\pbdotdot{{\bold {\,\ddot {\! \text{\it p}}}}}
\def\Pbdot{{\bold {\,\,\dot {\!\! \text{\it P}}}}}
\def\zetadot{\dot \zeta}

\def\[[{[\![}	\def\]]{]\!]}

\def\L{\bigl}	\def\R{\bigr}

\hyphenation{ge-o-des-ic ge-o-des-ics}

\documentstyle{amsppt}
\loadbold
\loadeusm
\TagsOnRight

\magnification=\magstephalf

\hoffset=.25truein
\hsize=6.0truein
\vsize=9.0truein
\voffset=-.25truein         % This is here because AMSTeX has started
                            % putting a 9 inch page 1/4 inch too low.
\parindent=20pt

\topmatter
\title Space-Time-\!-Time:  Five-Dimensional Kaluza--Weyl Space \endtitle
\author Homer G. Ellis \endauthor

\address
\vskip -30pt
\rm
$$
\alignat 2
\text{First version:  }& \text{Space-Time-\!-Time (1989)}& \\
\vspace{-3pt}
\text{Current version:  }& \hskip 49pt \text{April, 2001}& \\
\vspace{-3pt}
\text{Revised:  }& \hskip 53pt \text{July, 2001}&
\endalignat
$$
\vskip -5pt
\indent Homer G. Ellis\newline
\indent Department of Mathematics\newline
\indent University of Colorado at Boulder\newline
\indent 395 UCB\newline
\indent Boulder, Colorado  80309-0395\newline\newline
\indent Telephone: (303) 492-7754 (office); (303) 499-4027 (home)\newline
\indent Email:  ellis\@euclid.colorado.edu\newline
\indent Fax: (303) 492-7707
\endaddress

\abstract
\it Space-time-\!-time couples Kaluza's five-dimen\-sional geometry with Weyl's
conformal space-time geometry to produce an extension of space-time theory
that goes beyond what either of the Weyl and the Kaluza theories can achieve
by itself.  Kaluza's ``cylinder condition'' is replaced by an ``exponential
expansion constraint'' that causes translations along the secondary time
dimension to induce both the electromagnetic gauge transformations found in
the Kaluza and the Weyl theories and the metrical gauge transformations unique
to the Weyl theory, related exactly as Weyl had postulated.  A
space-time-\!-time geodesic describes a test particle whose rest mass $\mo$,
space-time momentum $\mo u^\mu$, and electric charge $q$, all defined
kinematically, evolve in accord with definite dynamical laws.  Its motion,
projected onto space-time, is governed by four apparent forces:  the Einstein
gravitational force, the Lorentz electromagnetic force, a force proportional
to the electromagnetic four-potential, and a force proportional to a scalar
field's gradient $d(\ln \phi)$.  The particle appears suddenly at an event
${\eusm E}_1$ with $q = - \phi ({\eusm E}_1)$ and disappears at an event
${\eusm E}_2$ with $q = \phi ({\eusm E}_2)$.  At ${\eusm E}_1$ and
${\eusm E}_2$ the gradient force infinitely dominates the others, causing
${\eusm E}_1$ and ${\eusm E}_2$ to occur preferentially in the valley depths of
the potential $\ln \phi$ --- this suggests the possibility of explaining some
aspects of atomic structure without invoking quantum theory.  Test particles
with $\mo = 0$ and $q \neq 0$ can exist, but must follow paths $p$ for which
$\phi (p) = \text{const}$, and must have $q = \pm \phi (p)$.  A
$\phi @,@,@,$-wave form with a null propagation vector can carry such
particles from place to place at the speed of light, keeping them immune to
the influence of an accompanying electromagnetic wave form.  Test particles
sharing a common event $\eusm E$ of appearance, with $q = - \phi (\eusm E)$,
or disappearance, with $q = \phi (\eusm E)$, can be made to ``interact'' by
demanding that the sum of their space-time-\!-time momenta at $\eusm E$
vanish.  This space-time-\!-time conservation law would comprise for such
interactions both conservation of space-time momentum and conservation of
electric charge.
\endabstract

\endtopmatter

\document

\baselineskip=15pt

\vskip -15pt

\line{\bf 1.  INTRODUCTION \hfil}
\vskip 3pt

\indent The theory I am going to describe here employs Kaluza`s
five-dimen\-sional geometry of 1919 \cite{1}, altered to a form that
encompasses Weyl's conformal space-time geometry of 1918 \cite{2}.  These two
early geometrical enlargements of Einstein's space-time theory of gravity to
include Maxwell's theory of electromagnetism were somewhat successful, each in
its own way, but they bore no apparent relation to one another.  Properly
joined, they make a theory, the theory of ``space-time-\!-time,'' that goes
well beyond what either is able to achieve by itself, and that differs
essentially from standard gauge theories of Kaluza--Klein type.

Weyl, to take into account the freedom to specify arbitrarily at each
space-time event a scale against which to measure the lengths of tangent
vectors at that event, enlarged the study of individual space-time metrics to
the study of whole families of conformally related space-time metrics.  He
postulated that transport of a tangent vector keeping its covariant derivative
equal to zero need not preserve its length with respect to any of these
metrics.  The consequent ``nonintegrability'' of lengths of vectors transported
in this manner around closed circuits he ascribed to inexactness of an
electromagnetic covector (1-form) potential $A$ whose exterior derivative
$d_\wedge @!@!@!@! A$ manifests as an electromagnetic field.  Conformal
transitions $G \to e^{2 \lambda} G$ between metrics coincided with transitions
$A \to A + d \lambda$ between potentials.  Weyl referred to invariance under
these combined transitions as ``gauge'' invariance, so the transitions have
come to be known as ``gauge transformations.''

Kaluza, taking a different tack, enlarged the study of space-time to the study
of five-dimen\-sional metric manifolds $\eusm M$ whose cross sections
transverse to the fifth dimension are space-time manifolds.  To account for the
unobservability of this extra dimension he postulated that translations of
$\eusm M$ in its direction should induce isometries of the metric $\Ghat$ of
$\eusm M$.  This condition, which he termed ``cylinder condition,'' can be
formulated as the requirement that there exist on $\eusm M$ a vector field
$\xi$, in the direction of the extra dimension, such that
${\eusm L}_\xi \Ghat = 0$, where ${\eusm L}_\xi$ denotes Lie differentiation
along~$\xi$.  The electromagnetic field grows out of nonintegrability of the
distribution of hyperplanes orthogonal to $\xi$, which traces back to
inexactness of a space-time electromagnetic covector potential~$\Ao$.
Transformations $\Ao \to \Ao + d \lambda$ leaving $d_\wedge @!@!@!@! \Ao$ and
therefore the electromagnetic field unchanged coincide with refoliations of
$\eusm M$ by space-time cross sections.

Space-time-\!-time theory brings together these seemingly disparate approaches
to the task of producing a unified theory of gravity and electromagnetism.  It
accomplishes this simply by replacing the isometry equation
${\eusm L}_\xi \Ghat = 0$ in Kaluza's cylinder condition by the conformality
equation ${\eusm L}_\xi \Ghat = 2 G$, where $G$ is the ``space-time part''
of~$\Ghat$.  This modification causes translations of $\eusm M$ along $\xi$ to
induce conformal transformations of the space-time metrics of the cross
sections of $\eusm M$ transverse to~$\xi$.  The result is a natural hybrid of
the Kaluza and the Weyl geometries that retains and enhances the most useful
characters of its parents while attenuating to benign and useful form those
that have caused difficulty.  Most notably, it retains both Kaluza's extra
dimension and Weyl's association of metrical with electromagnetic gauge
changes.  Also, it converts the objectionable nonintegrability of length
transference in the Weyl geometry to integrability without sacrificing the
principle that length, because it is a comparative measure, depends on
selection of a scale at each point, that is, on choice of a gauge.  In the
process it lends to the fifth dimension an essential significance that the
Kaluza geometry fails to provide.  This significance arises from a geometrical
construction that compels interpretation of the fifth dimension as a secondary
temporal dimension \cite{3}, in contrast to its more usual interpretation as a
spatial dimension whose unobservability has to be excused.

The picture that emerges from application of this hybrid geometry to the
modeling of physical systems has in it some surprising, unorthodox
representations of elementary physical phenomena, quantum phenomena included.
Taken on their own terms they offer the possibility of adding to our image of
the world a certain coherency not present in existing representations.  Whether
they are accurate will be, of course, a matter for investigation.

The geometry of space-time-\!-time is a special case of the geometry of what
may be called Kaluza--Weyl spaces, which conform to the requirement that
${\eusm L}_\xi \Ghat = 2 G$, but are unrestricted as to dimensionality of
the carrying manifold and signature of the metric. Sections 2--6 below present
the bare bones of this Kaluza--Weyl geometry, including a discussion of its
gauge transformations and ending with equations for its geodesics.  Section 7
develops the dynamics of test particles following space-time-\!-time geodesics.
Section 8 draws inferences about the behavior of these test particles and
proposes a conservation law for their interactions.  Section 9, the last, is
devoted to remarks speculative and prospective in nature.  A subsequent paper
will present field equations appropriate to the Kaluza--Weyl geometry.  

\vskip 12pt

\line{\bf 2.  KALUZA--WEYL SPACES \hfil}
\vskip 3pt

\indent Let $\eusm M$ be a manifold and $\Ghat$ a symmetric, nondegenerate
metric on~$\eusm M$.  The condition on $\eusm M$ and $\Ghat$ that will
replace Kaluza's ``cylinder condition'' as formulated in Sec\. 1 is the
\vskip 5pt

\parindent=30pt
{\narrower\smallskip
\noindent {\bf Exponential Expansion Constraint (EEC)}@:  There exists on
$\eusm M$ a vector field $\xi$ such that ${\eusm L}_\xi \Ghat = 2 G$, where
$G := \Ghat - \L(\Ghat \xi \xi\R)^{-1} \L(\Ghat \xi \otimes \Ghat \xi\R)$.
\smallskip}
\vskip 5pt
\parindent=20pt

\noindent When this constraint is satisfied let us call $\Ghat$ a
{\bf Kaluza--Weyl metric} and the pair $\{\eusm M, \Ghat\}$ a
{\bf Kaluza--Weyl space}.

For proper interpretation of the EEC the cotangent space $T_P$ of $\eusm M$ at
a point $P$ must be understood as the space of all linear mappings of the
tangent space $T^P$ into $\Bbb R$, and the tensor product $T_P \otimes T_P$ as
the space of all linear mappings of $T^P$ into $T_P$, one such being
$\Ghat (P)$.  This makes $\Ghat \xi$ a covector field on $\eusm M$
(the ``metric dual'' of $\xi$), and $\Ghat \xi \xi$ a scalar field on $\eusm M$
(the ``square length'' of $\xi$), whereupon $G$ is seen to be the orthogonal
projection of $\Ghat$ along~$\xi$, in that $G \xi = 0$ and $\Ghat v = G v$ if
$\Ghat \xi v = 0$.  Implicit in the EEC is that $\Ghat \xi \xi$ vanishes
nowhere, that, to put it differently, $\xi$ is nowhere null with respect to
$\Ghat$; a consequence is that $\xi$ itself vanishes nowhere.

A prototype for Kaluza--Weyl metrics is the de Sitter
space-time metric, which in the Lema\^\i tre coordinate system takes the form
$$
\Ghat = e^{2t} (dx \otimes dx + dy \otimes dy + dz \otimes dz)
          - R^2 (dt \otimes dt),
\tag1
$$
where $R$ is the (uniform) space-time radius of
curvature \cite{3, 4}.  Here
$ \xi = \partial / \partial t$, $\Ghat \xi = -R^2 dt$,
$ \Ghat \xi \xi = -R^2$, and $G = e^{2t} (dx \otimes dx +
dy \otimes dy + dz \otimes dz)$.

Space-time-\!-time metrics are those five-dimen\-sional Kaluza--Weyl metrics
$\Ghat$ for which $G$ has a space-time signature.  Prototypes are
the hyper-de Sitter metrics $\Ghat_\pm$ given by
$$
\Ghat_\pm
  = e^{2 \zeta} (dx \otimes dx + dy \otimes dy + dz \otimes dz - dt \otimes dt) 
        \pm R^2 (d \zeta \otimes d \zeta).
\tag2
$$
For both metrics $\xi = \partial / \partial \zeta$ and $G = e^{2 
\zeta} (dx \otimes dx + dy \otimes dy + dz \otimes dz - dt \otimes dt)$;
but $\Ghat_+ \xi \xi = R^2$, whereas $\Ghat_- \xi \xi = -R^2$, which of course
reflects the fact that $\Ghat_+$ has diagonal signature $+++-+$ and
$\Ghat_-$ has it $+++--$.  Like $\Ghat$ in Eq\. (1), each of $\Ghat_+$ and
$\Ghat_-$ gives to its carrying manifold $\eusm M$ a uniform radius of
curvature $R$.

If one removes the factor $e^{2 \zeta}$ from Eq\. (2), the resulting metrics
will satisfy Kaluza's isometry equation ${\eusm L}_\xi \Ghat = 0$, but the
ambiguity of signature will remain.  More generally, if $\Ghat$ satisfies
either the cylinder condition or the EEC, and $G$ has signature $+++\,\,@!@!-$
then $\Ghat$'s signature will be $+++-+$ or $+++--$, according as
$\Ghat \xi \xi > 0$ or $\Ghat \xi \xi < 0$.  Thinking it necessary to choose
between these signatures for $\Ghat$, Kaluza apparently opted for $+++-+$.
\footnote{
Kaluza was not committed to this choice, indeed seemed willing to let
it go the other way if by so doing he could overcome a ``serious difficulty''
pointed out to him by Einstein \cite{1, p\. 971}.}
As the first three +'s refer to spatial dimensions, one naturally is tempted
to say that this causes Kaluza's extra dimension to be spatial also.  But that
is mere verbal analogy --- it lacks any real justification in the form of a
conceptual parallelism between the fifth dimension, its coordinate generated
along $\xi$, and the three dimensions of physical space represented by the
first three coordinates.  Indeed, the geometric construction described in
\cite{3} makes it clear that the natural parallelism is with the fourth,
temporal dimension.  That parallelism is in fact on display here in the
similarity between the exponential role that $t$ plays in the de Sitter metric
and the exponential role that $\zeta$ plays in the hyper-de Sitter metrics.
Its existence is the reason why I attach the label {\bf space-time-\!-time} to
every five-dimen\-sional Kaluza--Weyl space $\{\eusm M, \Ghat\}$ for which $G$
has a space-time signature, irrespective of whether $\Ghat \xi \xi > 0$ or
$\Ghat \xi \xi < 0$.  There is, however, no implication that the secondary time
dimension is interchangeable with the primary.  The secondary is a child of the
primary, not a clone.

\vskip 12pt

\line{\bf 3.  CANONICAL FORMS OF KALUZA--WEYL METRICS \hfil}
\vskip 3pt

\indent Let $\{\eusm M, \Ghat\}$ be a Kaluza--Weyl space.  One sees easily that
$$
\Ghat = G + \epshat \phi^2 (A \otimes A),
\tag3
$$
where $\phi := \L| \Ghat \xi \xi \R|^{1/2}$,
$A := \L(\Ghat \xi \xi\R)^{-1} \Ghat \xi$, and
$\epshat := \text{sgn} \L(\Ghat \xi \xi\R) = 1$ or $-1$.  The projected metric
$G$, the scalar field $\phi$, and the covector field $A$ behave in the
following ways under Lie differentiation along $\xi$: ${\eusm L}_\xi \phi = 0$,
${\eusm L}_\xi A = 0$, and ${\eusm L}_\xi G = 2 G$.  This is demonstrable
by a few simple calculations.  First, $G \xi =
\Ghat \xi - \L(\Ghat \xi \xi\R)^{-1} \L(\Ghat \xi \otimes \Ghat \xi\R) \xi =
\Ghat \xi - \L(\Ghat \xi \xi\R)^{-1} \L(\Ghat \xi \xi\R) \Ghat \xi$, so
$G \xi = 0$ (as noted previously).
Next, because ${\eusm L}_\xi \xi = 0$, one has that
${\eusm L}_\xi \L(\Ghat \xi\R) = \L({\eusm L}_\xi \Ghat\R) \xi = 2 G \xi = 0$
and
${\eusm L}_\xi \L(\Ghat \xi \xi\R) = \L({\eusm L}_\xi \L(\Ghat \xi\R)\R) \xi
= 0$, so that clearly ${\eusm L}_\xi \phi = 0$ and ${\eusm L}_\xi A = 0$.
From Eq\. (3) it then follows that ${\eusm L}_\xi \Ghat = {\eusm L}_\xi G$,
whence ${\eusm L}_\xi G = 2 G$.

A decomposition of $G$ comes from integrating the differential equation
${\eusm L}_\xi G = 2 G$, the result being that $G = e^{2C} \Go$, where $C$ is a
scalar field, ${\eusm L}_\xi C = 1$, $\Go$ is a metric of the same signature as
$G$, and ${\eusm L}_\xi \Go = 0$.  Application of ${\eusm L}_\xi$ to both sides
of
$$
\Ghat = e^{2C} \Go + \epshat \phi^2 (A \otimes A)
\tag4
$$
then shows that this representation for $\Ghat$, under the conditions that
${\eusm L}_\xi C = 1$ and the Lie derivatives along $\xi$ of $\Go$, $\phi$, and
$A$ all vanish, is sufficient to make $\Ghat$ satisfy the EEC (with respect
to~$\xi$).  With these conditions the representation (4) therefore constitutes
a characterization of Kaluza--Weyl metrics.

Now let us introduce (by a standard construction) a coordinate system
$\[[ x^\mu , \zeta @,@,@,@, \]]$ adapted to $\xi$ so that
$\xi = \partial / \partial \zeta$.
\footnote{
Here $\mu$ and other Greek letter indices will range, if $d > 1$, from 1 to
$d - 1$, where $d := \text{dim}\;{\eusm M}$; if $d = 1$, then the only
coordinate is $\zeta$, so $\mu$ does not enter the picture.  $M$ and other
upper case roman indices will range from 1 to d.}
As a covector field, $A$ has in $\[[ x^\mu , \zeta @,@,@,@, \]]$ the expansion
$A = A_\mu dx^\mu + A_\zeta d \zeta$.  But $A_\zeta =
A (\partial / \partial \zeta) = A \xi = (\Ghat \xi \xi)^{-1} \Ghat \xi \xi =
1$, so $A = A_\mu dx^\mu + d \zeta$.  Further, $0 = {\eusm L}_\xi A =
{\eusm L}_{\partial / \partial \zeta} A =
(\partial A_\mu / \partial \zeta) dx^\mu$, so
$\partial A_\mu / \partial \zeta = 0$; thus the $A_\mu$ depend on the
coordinates $x^\kappa$ alone, and not on~$\zeta$.  Also,
$\partial \phi / \partial \zeta = {\eusm L}_\xi \phi = 0$, so $\phi$ is a
function of the $x^\kappa$ only.  Arguing similarly about $G$, we arrive at the
adapted coordinates version of Eq\. (3), viz.,
$$
\Ghat = dx^\mu \otimes g_{\mu \nu} dx^\nu
         + \epshat \phi^2 (A_\mu dx^\mu + d \zeta)
                  \otimes (A_\nu dx^\nu + d \zeta),
\tag"($3'$)"
$$
with $\partial \phi / \partial \zeta = \partial A_\mu / \partial \zeta = 0$
and $\partial g_{\mu \nu} / \partial \zeta = 2 g_{\mu \nu}$.  To do the same
for Eq\. (4), let us specify the scalar field $C$, which is at our disposal.
The possibilities are $C = \zeta  + \theta$, thus
$e^{2C} = e^{2 \zeta} e^{2 \theta}$, with
$\partial \theta / \partial \zeta = 0$.  Inasmuch as the factor $e^{2 \theta}$
can be absorbed into $\Go$, we can, without loss of generality, take
$\theta = 0$.  Then $G = e^{2 \zeta} \Go$ and
$g_{\mu \nu} = e^{2 \zeta} \go_{\mu \nu}$, with
$\partial \go_{\mu \nu} / \partial \zeta = 0$.  Let us also introduce the
covector field $\Ao := A - d \zeta$, for which $A = \Ao + d \zeta$, $\Ao =
\Ao_\mu dx^\mu$, $\Ao_\mu = A_\mu$, ${\eusm L}_\xi \Ao = 0$, and
$\partial \Ao_\mu / \partial \zeta = 0$.  Then Eq\. (4) takes the forms
$$
\align
\Ghat
 &= e^{2 \zeta} \Go + \epshat \phi^2 \L(\Ao + d \zeta\R)
                             \otimes \L(\Ao + d \zeta\R) \\
\vspace{-7pt}
\tag"($4'$)" \\
\vspace{-7pt}
 &= e^{2 \zeta} \L(dx^\mu \otimes \go_{\mu \nu} dx^\nu\R) +
        \epshat \phi^2 \L(\Ao_\mu dx^\mu + d \zeta\R) \otimes
        \L(\Ao_\nu dx^\nu + d \zeta\R),
\endalign
$$
with $\partial \phi / \partial \zeta = \partial \Ao_\mu / \partial \zeta =
\partial \go_{\mu \nu} / \partial \zeta = 0$.  These forms are canonical for
Kaluza--Weyl metrics.  They differ from the analogous canonical forms for
metrics satisfying Kaluza's cylinder condition precisely by the presence of
the factor $e^{2 \zeta}$.  This factor produces surprising effects, as we shall
see.

\vskip 12pt

\line{\bf 4.  GAUGE TRANSFORMATIONS \hfil}
\vskip 3pt

\indent When $\{\eusm M, \Ghat\}$ is a space-time-\!-time, the tensor field
$F @!@! := @!@! -2 d_\wedge @!@!@!@! A$ will come to be identified as the
electromagnetic field tensor.  We shall have then that 
$F = -2 d_\wedge @!@!@!@! \L(\Ao + d \zeta\R) = -2 d_\wedge @!@!@!@! \Ao$,
thus that $\Ao$ takes the role of electromagnetic four-covector potential.
Klein, who independently formulated and refined the Kaluza geometry \cite{5},
and Einstein, who introduced refinements of his own \cite{6}, used the same
identification of $F$ for the Kaluza (--Klein) theory, deviating somewhat from
Kaluza's choice.  They further recognized that the electromagnetic gauge
transformations $\Ao \to \Ao' := \Ao + d \lambda$ such that
${\eusm L}_\xi \lambda = 0$ are generated by transformations
$\zeta \to \zeta' := \zeta - \lambda$ such that
$\partial \lambda / \partial \zeta = 0$, which follows from
$A = \Ao' + d \zeta' = \Ao + d \zeta$ and $d_\wedge @!@!@!@! A =
d_\wedge @!@!@!@! \Ao' = d_\wedge @!@!@!@! \Ao$.  This recognition was the
first step on the road to the gauge theories that now abound in theoretical
physics.  Missing from Kaluza--Klein theory and from these later gauge
theories, however, is any remembrance of Weyl's earlier association of
electromagnetic gauge transformations with (conformal) gauge transformations
of the metric of space-time.
\footnote{
Weyl himself contributed to this amnesia by transferring his allegiance over to
an association of electromagnetic gauge transformations with electron wave
field phase shifts \cite{7}, the association that London extracted from the
theories of Weyl and of Kaluza and Klein \cite{8}.}
In space-time-\!-time this association is preserved through intermediation of
the coordinate transformation $\zeta' = \zeta - \lambda$, for the conformality
relation $\Go' = e^{2 \lambda} \Go$ is a clear implication of
$G = e^{2 \zeta} \Go = e^{2 \zeta'} \Go'$.

The coordinate transformations that generate the electromagnetic and the
metrical gauge transformations, {\it being} coordinate transformations, do
not alter the metric of space-time-\!-time.  This is a principal advantage that
the space-time-\!-time geometry has over the Weyl geometry.  Weyl, working 
without the aid of a fifth dimension, impressed his infinitude of conformally
related space-time metrics onto one four-dimensional manifold.  That is very
much like drawing all the maps of the world on a single sheet of paper, a
practice that would conserve paper but confound navigators.  In effect, the
space-time-\!-time geometry economizes on paper but avoids the confusion of
maps on maps, by drawing a selection of the maps on individual sheets, then
stacking the sheets so that each of the remaining maps can be generated on
command by slicing through the stack in a particular way.  The Kaluza--Klein
geometry does much the same, but the cylinder condition restricts its stack
to multiple copies of a single map, with no new maps producible by slicing.

\vskip 12pt

\line{\bf 5.  CONNECTION FORMS AND COVARIANT DIFFERENTIATIONS \hfil}
\vskip 3pt

\indent A coframe system $\{\omega^\mu, \omega^d\}$ that will facilitate
computation of connection forms for the Kaluza--Weyl space $\{\eusm M, \Ghat\}$
is defined as follows:  relabeling the coordinates $x^\mu$ as $x^{\mu'}$, let 
$\omega^\mu := dx^{\mu'} J_{\mu'} {}^\mu$, with $[J_{\mu'} {}^\mu ]$ and its
inverse matrix $[J_{\mu} {}^{\mu'} ]$ independent of $\zeta$; let
$\omega^d := \phi A$.  In this system $\Go$ has the expansion
$\Go = \omega^\mu \otimes \go_{\mu \nu} \omega^\nu$, where $\go_{\mu \nu} =
J_\mu {}^{\mu'} \go_{\mu' \nu'} J_\nu {}^{\nu'}$, and $\Ghat$ takes the
semi-orthogonal form
$$
\Ghat = e^{2 \zeta} \L(\omega^\mu \otimes \go_{\mu \nu} \omega^\nu\R)
            + \epshat \L(\omega^d \otimes \omega^d\R). 
\tag5
$$
Upon identifying the frame system $\{ e_\mu , e_d \}$ to which
$\{ \omega^\mu , \omega^d \}$ is dual, one has
$$
\align
e_\mu
 &= J_\mu {}^{\mu'} \L(\partial_{\mu'} - \Ao_{\mu'} \partial_\zeta\R)
    = \partial_\mu - \Ao_\mu \partial_\zeta, \\
\vspace{-7pt}
\tag6 \\
\vspace{-7pt}
e_d
 &= \phi^{-1} \xi = \phi^{-1} \partial_\zeta, \\
\vspace{-7pt}
\intertext{to go with}
\vspace{-7pt}
\omega^\mu
 &= dx^{\mu'} J_{\mu'} {}^\mu, \\
\vspace{-7pt}
\tag7 \\
\vspace{-7pt}
\omega^d
 &= \phi A = \phi \L(\Ao_{\mu'} dx^{\mu'} + d \zeta\R)
           = \phi \L(\Ao_\mu \omega^\mu + d \zeta\R),
\endalign
$$
where $\partial_{\mu'} := \partial / \partial x^{\mu'}$, $\partial_\zeta :=
\partial / \partial \zeta$, $\partial_\mu :=
J_{\mu} {}^{\mu'} \partial_{\mu'}$, and $\Ao_\mu =
J_{\mu} {}^{\mu'} \Ao_{\mu'}$, so that $\Ao = \Ao_\mu \omega^\mu$.
The components $\go_{\mu \nu}$ and $\Ao_\mu$ are independent of~$\zeta$.
The vector field $e_d$ is the unit
\linebreak
normalization of $\xi$ and is
orthogonal to each of the vector fields~$e_\mu$.  It is not difficult
to see that ${\eusm L}_\xi e_\mu = {\eusm L}_\xi e_d = 0$  and
${\eusm L}_\xi \omega^\mu = {\eusm L}_\xi \omega^d = 0$.  Thus
we have a frame system and its dual coframe system that are Lie constant 
along $\xi$, but with the further property that $e_d$ has
unit length and is orthogonal to each~$e_\mu$.  Their constancy
along $\xi$ makes them gauge invariant:  coordinate gauge
transformations $\zeta \to \zeta - \lambda$ leave them unchanged.
\footnote{
In the terminology of fibre bundle theory the $e_\mu$ and the tangent
subspace they span at a point are ``horizontal,'' and $e_d$ and the subspace
it spans at a point are ``vertical,'' as determined with reference to the
covector field $A$, standing in for a bundle connection 1-form.}

For the exterior derivatives of $\omega^\mu$ and $\omega^d$ we have
$$
\align
d_\wedge @!@!@!@! \omega^\mu
 &= C_\kappa {}^\mu {}_\lambda \omega^\lambda \wedge \omega^\kappa,
\tag8 \\
\vspace{-9pt}
\intertext{and}
\vspace{-9pt}
d_\wedge @!@!@!@! \omega^d
 &= - (1/2) \phi F + d \phi \wedge A
\vspace{-7pt}
\tag9 \\
\vspace{-7pt}
 &= - (1/2) \phi F_{\kappa \lambda} \omega^\lambda \wedge \omega^\kappa
       + \phi^{-1} \phi_{. \lambda} \omega^\lambda \wedge \omega^d,
\endalign
$$
with $C_\kappa {}^\mu {}_\lambda$ skew-symmetric in $\kappa$ and
$\lambda$ and independent of $\zeta$, and with
$$
\align
F :\!
 &= - 2d_\wedge A = - 2d_\wedge \Ao
    = F_{\kappa \lambda} \omega^\lambda \wedge \omega^\kappa,
\tag10 \\
\vspace{-9pt}
\intertext{where}
\vspace{-9pt}
F_{\kappa \lambda}
 &= \Ao_{\lambda . \kappa}
     - \Ao_{\kappa . \lambda}
     - 2 \Ao_\mu C_\kappa {}^\mu {}_\lambda,
\tag11
\endalign
$$ 
also skew-symmetric in $\kappa$ and $\lambda$ and independent of~$\zeta$.
Here $f_{. \mu} := \partial_\mu f$, for scalar fields $f$.

The torsionless covariant differentiation $\bdhat$ on $\eusm M$ that is
compatible with $\Ghat$ has connection forms $\omegahat_K {}^M$ such that
$\bdhat e_K = \omegahat_K {}^M \otimes e_M$ and $\bdhat @,@, \omega^M  =
- \omegahat_K {}^M \otimes \omega^K$.
They can be expressed as follows:
$$
\align
\omegahat_\kappa {}^\mu
 &= \omega_\kappa {}^\mu
     + \L[\phi^{-1} \go_\kappa {}^\mu
     + \epshat (1/2) e^{-2 \zeta} \phi \Fo_\kappa {}^\mu\R] \omega^d, \\
\omegahat_\kappa {}^d
 &= - \L[\epshat e^{2 \zeta} \phi^{-1} \go_{\kappa \lambda}
                   - (1/2) \phi F_{\kappa \lambda}\R] \omega^\lambda
        + \phi^{-1} \phi_{. \kappa} \omega^d, \\
\omegahat_d {}^\mu
 &= \L[\phi^{-1} \go^\mu {}_\lambda
         - \epshat (1/2) e^{-2 \zeta} \phi \Fo^\mu {}_\lambda\R] \omega^\lambda
      - \epshat e^{-2 \zeta} \phi^{-1} \phio^{. \mu} \omega^d,
\tag12 \\ 
\vspace{-9pt}
\vspace{-7pt}
\intertext{and}
\vspace{-9pt}
\omegahat_d {}^d
 &= 0.
\endalign
$$
In these and in subsequent equations raising of an index with $\go^{\mu \nu}$
is indicated by insertion of a $\; \degree \;$, unless one is already
present, as in $\go_\kappa {}^\mu := \go_{\kappa \lambda} \go^{\lambda \mu}
= \delta_\kappa {}^\mu$ and $\Ao^\mu := \Ao_\lambda \go^{\lambda \mu}$.  Also,
$$
\align
\!\!\!\!\!\!\!\!\!\!\!
\omega_\kappa {}^\mu :\!
 &= \L(\Gammao_\kappa {}^\mu {}_\lambda
        + \Deltao_\kappa {}^\mu {}_\lambda\R) \omega^\lambda,
\tag13 \\
\vspace{-7pt}
\intertext{where}
\vspace{-7pt}
\!\!\!\!\!\!\!\!\!\!\!
\Gammao_\kappa {}^\mu {}_\lambda :\!
 &= (1/2) \L(\go_{\nu \lambda . \kappa}
              + \go_{\kappa \nu . \lambda}
              - \go_{\kappa \lambda . \nu}\R) \go^{\nu \mu} 
            - \L(C_\kappa {}^\mu {}_\lambda
                 + \Co_{\kappa \lambda} {}^\mu
                 + \Co_{\lambda \kappa} {}^\mu\R)
\tag14 \\
\vspace{-7pt}
\intertext{and}
\vspace{-7pt}
\!\!\!\!\!\!\!\!\!\!\!
\Deltao_\kappa {}^\mu {}_\lambda :\!
&= - \L(\Ao_\kappa \go^\mu {}_\lambda
         + \go_\kappa {}^\mu \Ao_\lambda
         - \go_{\kappa \lambda} \Ao^\mu\R),
\tag15 \\
\endalign
$$
with $\Co_{\kappa \lambda} {}^\mu := \go_{\lambda \nu} C_\kappa {}^\nu {}_\pi
\go^{\pi \mu}$.

A covariant differentiation $\bold d$ on $\eusm M$, related to but distinct
from $\bdhat$, is fixed by the stipulations that
$\bold d e_\kappa = \omega_\kappa {}^\mu \otimes e_ \mu$ and $\bold d e_d = 0$,
or, equally well, by $\bold d @,@, \omega^\mu =
- \omega_\kappa {}^\mu \otimes \omega^\kappa$ and $\bold d @,@, \omega^d = 0$.
This is a direct analog of the covariant differentiation in Weyl's geometry,
as it satisfies $\bold d G = 2 A \otimes G$, the principal characterizing
condition of Weyl's affine connection.  Although $\bold d$ is not in general
torsionless, $\text{Tor\ } \bold d = d_\wedge @!@!@!@! \omega^d \otimes e_d =
\L[-(1/2) F + \phi^{-1} d \phi \wedge @!@!@!@! A\R] \otimes~\xi$, so that the
components of torsion in directions orthogonal to $\xi$ vanish:
$\omega^\mu (\text{Tor\ } \bold d) =
(d_\wedge @!@!@!@! \omega^d) (\omega^\mu e_d) = 0$.

\vskip 12pt

\line{\bf 6.  GEODESIC EQUATIONS \hfil}
\vskip 3pt

\indent Let $p\: I \to { \eusm M }$ be a path in $\eusm M$, with parameter
interval $I$, and let the components of its velocity $\dot p$ be
$\{ {\dot p}^\mu , {\dot p}^d \}$, in the frame system $\{ e_\mu , e_d \}$.
For the acceleration of $p$ generated by the covariant differentiation
$\bdhat$ one has $\pbdotdot =
\pbdotdot@,@,@,@,@,^\mu e_\mu (p) + \pbdotdot@,@,@,@,@,^d e_d (p)$, where
$$
\align
\pbdotdot@,@,@,@,@,^\mu 
 &= ({\dot p}^\mu)\dot{}\,
     + {\dot p}^\kappa \omegahat_\kappa {}^\mu (p) {\dot p}
     + {\dot p}^d \omegahat_d {}^\mu (p) {\dot p}
\tag16 \\
\vspace{-7pt}
\intertext{and}
\vspace{-7pt}
\pbdotdot@,@,@,@,@,^d 
 &= \L({\dot p}^d\R)\dot{}\,
     + {\dot p}^\kappa \omegahat_\kappa {}^d (p) {\dot p}
     + {\dot p}^d \omegahat_d {}^d (p) {\dot p}.
\tag17
\endalign
$$
The condition that $p$ be an affinely parametrized geodesic path of
$\bdhat$ is that $\pbdotdot = 0$, which is equivalent to
$\pbdotdot@,@,@,@,@,^\mu = 0$ and $\pbdotdot@,@,@,@,@,^d = 0$.  These are
equivalent, respectively, to
$$
\align
\!\!\!\!\!\!\!\!\!\!\!\!\!\!\!\!
\L(e^{2 \zeta} {\dot p}^\mu\R)\dot{}\,
  + e^{2 \zeta} {\dot p}^\kappa \Gammao_\kappa {}^\mu {}_\lambda
                {\dot p}^\lambda
 &= \epshat \phi {\dot p}^d \Fo^\mu {}_\lambda {\dot p}^\lambda
    - e^{2 \zeta} {\dot p}^\kappa \go_{\kappa \lambda} {\dot p}^\lambda \Ao^\mu
    + \epshat {\dot p}^d {\dot p}^d \phi^{-1} \phio^{. \mu}
\!\!\!\!
\tag18 \\
\vspace{-7pt}
\intertext{and}
\vspace{-7pt}
\L(\epshat \phi {\dot p}^d\R)\dot{}\,
 &= e^{2 \zeta} {\dot p}^\kappa \go_{\kappa \lambda} {\dot p}^\lambda,
\tag19
\endalign
$$
in which for brevity the compositions with $p$ of the various scalar fields
are implicit rather than express.

As one knows, $\pbdotdot = 0$ implies that $\L[\Ghat (p) {\dot p}
{\dot p}\R]\,\dot{} = 0$, thus that $\Ghat (p) {\dot p} {\dot p}$ is
constant.  This takes the form
$$
e^{2 \zeta} {\dot p}^\kappa \go_{\kappa \lambda} {\dot p}^\lambda
    + \epshat {\dot p}^d {\dot p}^d = \epsilon,
\tag20
$$
where $\epsilon := \text{sgn} \L(\Ghat (p) {\dot p} {\dot p}\R) = 1$, 0, or
$-1$, provided that the parametrization of $p$ is by arc length when
$\Ghat (p) {\dot p} {\dot p} \neq 0$.

\vskip 12pt

\line{\bf 7.  TEST PARTICLE DYNAMICS IN SPACE-TIME-\!-TIME \hfil}
\vskip 3pt

\indent When the Kaluza--Weyl space $\{\eusm M, \Ghat\}$ is a
space-time-\!-time, its geodesics can be interpreted as histories of test
particles, just as is done with space-time geodesics.  It then becomes of
interest to learn the dynamics governing the motions of such test particles.
These dynamics will be, of course, only the kinematics imposed on the test
particles by the space-time-\!-time geometry, but dressed up in labels such
as momentum, mass, charge, and force.  For space-time test particles the
procedure is relatively straightforward, geodesics in space-time having no
kinematical variables to be interpreted as mass or electric charge, and only
the gravitational force to be assigned a kinematical identity.  In
space-time-\!-time there is a great deal more to be interpreted than in
space-time.  To arrive at useful interpretations we are bound to rely on
formal similarities with extant equations and concepts, but we must accept
whatever dynamics the kinematics dictate, and {\it firmly repress} the natural
tendency to insist upon complete agreement with preconceived notions of
particle properties and behavior derived from theories based on other, more
restrictive geometries, or on no geometry at all.

It will be convenient to have the signature of the space-time part of the
metric be $---\,+$; this causes the signature of $\Ghat$ to be $---++$ if
$\epshat = 1$, and $---+-$ if $\epshat = - 1$.
To begin, let us define the {\bf space-time-\!-time momentum covector} $P$ of
the test particle following the geodesic path $p$ to be the metric dual of its
velocity, that is, $P := \Ghat (p) {\dot p}$.  Because $\Ghat$ is
$\bdhat $-covariantly constant, $\Pbdot = \Ghat (p) \pbdotdot$, and therefore
the geodesic equation $\pbdotdot = 0$ is equivalent to $\Pbdot = 0$.
Analysis of the latter equation will yield the desired interpretations.

In the adapted coframe system $\{ \omega^\mu , \omega^d \}$ the momentum $P$
has the expansion $P = P_\kappa \omega^\kappa (p) + P_d \omega^d (p)$, where
$$
\align
P_\kappa
 &= e^{2 \zeta} {\dot p}^\mu \go_{\mu \kappa}
\tag21 \\
\vspace{-7pt}
\intertext{and}
\vspace{-7pt}
P_d
 &= \epshat {\dot p}^d.
\tag22 \\
\endalign
$$
The covariant derivative of $P$ has the expansion $\Pbdot =
\Pbdot_\kappa \omega^\kappa (p) + \Pbdot_d \omega^d (p)$, where
$$
\align
\Pbdot_\kappa
 &= (P_\kappa)\dot{}\,
     - P_\mu \omegahat_\kappa {}^\mu (p) {\dot p}
     - P_d \omegahat_\kappa {}^d (p) {\dot p} \\
\vspace{-7pt}
\tag23 \\
\vspace{-7pt}
 &= (P_\kappa)\dot{}\,
     - P_\mu \Gammao_\kappa {}^\mu {}_\lambda {\dot p}^\lambda
     - \phi P_d F_{\kappa \lambda} {\dot p}^\lambda
     + e^{2 \zeta} {\dot p}^\mu \go_{\mu \nu} {\dot p}^\nu \Ao_\kappa
     - \epshat P_d P_d \phi^{-1} \phi_{. \kappa}
\vspace{-7pt}
\intertext{and}
\vspace{-7pt}
\Pbdot_d
 &= (P_d)\dot{}\,
     - P_\mu \omegahat_d {}^\mu (p) {\dot p}
     - P_d \omegahat_d {}^d (p) {\dot p} \\
\vspace{-7pt}
\tag24 \\
\vspace{-7pt}
 &= (P_d)\dot{}\,
     + P_d \phi^{-1} \phi_{. \mu} {\dot p}^\mu
     - e^{-2 \zeta} \phi^{-1} P_\mu \go^{\mu \nu} P_\nu.
\endalign
$$
Let
$$
\align
\mo :\!
 &= (P_\mu \go^{\mu \nu} P_\nu)^{1/2}
    = e^{2 \zeta} ({\dot p}^\mu \go_{\mu \nu} {\dot p}^\nu)^{1/2}
\tag25 \\ 
\vspace{-7pt}
\intertext{and}
\vspace{-7pt}
q :\!
 &= P \xi (p) = \phi P_d = \epshat \phi {\dot p}^d \\
\vspace{-7pt}
\tag26 \\
\vspace{-7pt}
 &= \epshat \phi^2 A(p) {\dot p}
    = \epshat \phi^2 \L(\Ao_\mu {\dot p}^\mu + \zetadot\R).  
\endalign
$$
In terms of these Eq\. (20) becomes
$$
e^{-2 \zeta} \mo^2 + \epshat (q / \phi)^2 = \epsilon,
\tag27
$$
and the equations $\Pbdot_\kappa = 0$ and $\Pbdot_d = 0$, equivalent
to $\Pbdot = 0$, are seen to be further equivalent to
$$
\align
(P_\kappa)\dot{}\,
 &= P_\mu \Gammao_\kappa {}^\mu {}_\lambda {\dot p}^\lambda
     + q F_{\kappa \lambda} {\dot p}^\lambda
     - e^{-2 \zeta} \mo^2 \Ao_\kappa
     + \epshat (q/\phi)^2 \phi^{-1} \phi_{. \kappa} \\
\vspace{-7pt}
\tag28 \\
\vspace{-7pt}
 &= e^{-2 \zeta} \L(P_\mu \Gammao_\kappa {}^\mu {}^\lambda P_\lambda
      + q \Fo_\kappa {}^\lambda P_\lambda
      -  \mo^2 \Ao_\kappa\R)
      + \epshat (q/\phi)^2 \phi^{-1} \phi_{. \kappa} \\
\vspace{-7pt}
\intertext{and}
\vspace{-7pt}
\dot q 
 &= e^{-2 \zeta} \mo^2.
\tag29
\endalign
$$
Equation (29) can be recast in light of Eq\. (27) as
$$
\align
\dot q
 &= \epsilon - \epshat (q / \phi)^2.
\tag30 \\
\vspace{-2pt}
\intertext{Together with Eqs\. (26) and (27) it implies that}
\vspace{-2pt}
\L(\mo^2\R)\dot{}\,
 &= 2 \L[-\mo^2 \Ao_\kappa + \epshat e^{2 \zeta} (q/\phi)^2
                             \phi^{-1} \phi_{. \kappa}\R] {\dot p}^\kappa.
\tag31
\endalign
$$

The scalar $\Go (p) {\dot p} {\dot p}$, recognizable also as
${\dot p}^\mu \go_{\mu \nu} {\dot p}^\nu$ and as
$e^{-4 \zeta} \mo^2$, may be positive, zero, or negative on
different geodesics and, generally, on different portions of the same geodesic.
It is the square length of the projection ${\dot p}^\mu e_\mu (p)$ along $\xi$
of the velocity ${\dot p}$, as measured by the space-time metric $\Go$ of
signature  $---\,+$.  Wherever on $p$ this scalar is positive, that is,
wherever the space-time projection of ${\dot p}$ is timelike, we can introduce
a proper-(primary)time  parameter $\tauo$ such that
$$
(\tauo)\dot{}\, = \L(\Go (p) {\dot p} {\dot p}\R)^{1/2}
                = ({\dot p}^\mu \go_{\mu \nu} {\dot p}^\nu)^{1/2}
                = e^{-2 \zeta} \mo,
\tag32
$$
and with it define space-time velocity components $u^\lambda$ by
$u^\lambda := {\dot p}^\lambda / (\tauo)\dot{}\,$.  Then Eqs\. (28), (29),
and (31) can transmute to
$$
\align
\frac {d P_\kappa}{d \tauo}
 &= P_\mu \Gammao_\kappa {}^\mu {}_\lambda u^\lambda
     + q F_{\kappa \lambda} u^\lambda
     - \mo \Ao_\kappa
     + \epshat e^{2 \zeta} \frac {(q/\phi)^2}{\mo} \phi^{-1} \phi_{. \kappa},
\tag33 \\
\vspace{5pt}
\frac {dq}{d \tauo}
 &= \mo,
\tag34 \\
\endalign
$$
and
$$
\frac {d \mo}{d \tauo} = \Bigl[-\mo \Ao_\kappa
                                + \epshat e^{2 \zeta} \frac {(q/\phi)^2}{\mo}
                                  \phi^{-1} \phi_{. \kappa}\Bigr] u^\kappa.
\tag35
$$
Equations (33) and (34) are coupled equations of motion for the test particle;
they have the subsidiary equation (35) as a consequence.

Finally, recall that by convention $P_\nu \go^{\nu \mu} =: \Po^\mu$.  From
this follows that
$\Po^\mu = e^{2 \zeta} {\dot p}^\mu$,
$\mo = \L(\Po^\mu \go_{\mu \nu} \Po^\nu\R)^{1/2}$,
and, wherever $\mo^2 > 0$, $\Po^\mu = \mo u^\mu$ and we may replace
the covector equation (33) by the equivalent vector equation
$$
\frac {d(\mo u^\mu)}{d \tauo}
  + (\mo u^\kappa) \Gammao_\kappa {}^\mu {}_\lambda u^\lambda
       = q \Fo^\mu {}_\lambda u^\lambda - \mo \Ao^\mu
          + \epshat e^{2 \zeta} \frac {(q/\phi)^2} \mo \phi^{-1} \phio^{. \mu},
\tag"($33'$)"
$$
which is a reformulation in present terms of the geodesic equation (18).

Comparison of Eqs\. (33) and ($33'$) with the classical relativistic equations
of motion for an electrically charged particle makes credible the
interpretation of $\mo$ as {\bf rest mass} and $q$ as {\bf electric charge}
of the test particle, of $P_\mu$ as components of its
{\bf space-time momentum covector},
and of $\mo u^\mu$ as components of its
{\bf space-time momentum vector}.  These interpretations adopted,
four ``forces'' appear in Eqs\. (33) and ($33'$) as drivers of the
momentum rates $d P_\kappa / d \tauo$ and $d (\mo u^\mu) / d \tauo$:
\vskip 5pt

\parindent=10pt
{\narrower\smallskip
\item{1.}
the Einstein force attributable to the gravitational field and other
space-time geometry fields, as manifested in the connection coefficients
$\Gammao_\kappa {}^\mu {}_\lambda$;
\item{2.}
the Lorentz force attributable to the electromagnetic field manifested
in the tensor~$F$;
\item{3.}
a force proportional to the electromagnetic potential field embodied in
the covector~$\Ao$;
\item{4.}
a force proportional to the scalar field gradient $d(\ln \phi)$
($= \phi^{-1} d\phi = \phi^{-1} \phi_{. \kappa} \omega^{\kappa}$).
\smallskip}
\vskip 3pt
\parindent=20pt

\noindent Of these only the first is present in space-time theory as part
of the geometry.  The first
and the second show up in the geometry of
Kaluza--Klein theory, and the fourth would
as well but for the restriction that
$\Ghat \xi \xi$, and therefore $\phi$, is constant, which Klein \cite{5} and
Einstein \cite{6} added to Kaluza's cylinder condition.
\footnote{
Einstein called the result ``sharpened cylinder condition.''}
Versions of the fourth occur in Kaluza's paper \cite{1, Eq\. (11a)} and,
implicitly, in Jordan's elaboration of Kaluza's theory set out by
Bergmann \cite{9}.  The third cannot be found in any of those theories --- its
existence here is owed specifically to the inclusion of Weyl's geometry in
space-time-\!-time theory by way of the EEC.  As a force that will bend the
tracks of particles in regions where the electromagnetic field $F$ vanishes
but the electromagnetic potential $\Ao$ does not, its presence offers the
chance of a new perspective on electron optics, on the Aharonov--Bohm effect
in particular \cite{10, 11}.
 
In space-time theory rest mass and electric charge are extraneous to the
geometry and are therefore of necessity put into equations of motion by hand,
usually as constants.
In space-time-\!-time $\mo$ and $q$ are kinematical variables of geometrical
origin which remain constant only in special cases --- witness Eq\. (29),
which allows $q$ to be constant only if $\mo = 0$, and Eq\. (31),
which requires a delicate balance if $\mo$ is not to vary.  The inconstancy
of $q$ stands in marked contrast to $q$'s behavior in Kaluza (--Klein) theory,
where instead of Eq\. (29) one encounters $\dot q = 0$, which makes $q$ a
constant of the motion.  Likewise, in Kaluza--Klein theory in place of
Eq\. (31) one has $\L(\mo^2\R)\dot{}\, = 0$, which makes $\mo$ a constant.  But in
Kaluza (--Jordan) theory Eq\. (31) is replaced by $\L(\mo^2\R)\dot{}\, =
2 \epshat (q/\phi)^2 \phi^{-1} \phi_{. \kappa} {\dot p}^\kappa$, which permits
$\mo$ to vary.  Unlike $q$, which has the gauge-invariant definition
$q := P \xi(p)$, $\mo$ as here defined is not gauge-invariant (although
vanishing of $\mo$ is). Its variability therefore poses fewer questions than
does that of $q$, and these can safely be excluded from the present discussion.
The issues raised by the variability of $q$, on the other hand, are of
considerable import.  They will be addressed in the next section.

\vskip 12pt

\line{\bf 8.  TEST PARTICLE BEHAVIOR IN SPACE-TIME-\!-TIME \hfil}
\vskip 3pt

\indent As a vacuum metric for space-time-\!-time we can reasonably choose
either of the hyper-de Sitter metrics of Eq\. (2), expecting test particle
behavior to differ somewhat between them.  It will be instructive to study
$\Ghat_-$, taken in the form $\Ghat = -\Ghat_-$ so that the signature of
$\Ghat$ will be $---++$ and the space-time signature will be
$---\,\,@!@!@!@!@!+$ as in the preceding section.  Then $\epshat = 1$,
$\phi = R$ (the uniform space-time-\!-time radius of curvature), and
$\Gammao_\kappa {}^\mu {}_\lambda$, $\Ao_\kappa$,
$F_{\kappa \lambda}$, and $\phi_{. \kappa}$ are all zero.

Equation (28) is simply $(P_\kappa)\dot{}\, = 0$, from which
$\[[ P_\kappa \]] = \[[ {-a}, -b, -c, E @,@,@,@, \]]$, a constant.  With
Eq\. (21) this implies that
$$
\[[ \dot x, \dot y, \dot z, \dot t @,@,@,@, \]]
   = \[[ a, b, c, E @,@,@,@, \]] e^{-2 \zeta},
\tag36
$$
which in turn implies, when $E \neq 0$, that $\bold v :=
\[[ dx/dt, dy/dt, dz/dt @,@,@,@, \]] = \[[ a/E, b/E, c/E @,@,@,@, \]]$,
thus that the motion is uniform in space-time.  The rest mass evolution
equation (31) is $(\mo^2)\dot{}\, = 0$, so $\mo$ is constant; in fact
$\mo := (P_\mu \go^{\mu \nu} P_\nu)^{1/2} = \sqrt{E^2 - a^2 - b^2 - c^2}$,
and therefore $\mo = |E| \sqrt{1 - |\bold v|^2}$, a familiar relation.
For a test particle traveling slower than light, $\mo^2 > 0$ and
$\epsilon = 1$, so the charge evolution equation (30) is
$\dot q = 1 - (q / \phi)^2$.  A solution of this equation representative of
those consistent with $\mo^2 > 0$ is
$$
q = \phi \tanh(\tauhat/\phi),
\tag37
$$
where $\tauhat$ is an arc length parameter for the test particle's geodesic.
Equations (26) and (37) imply that
$\dot \zeta = \phi^{-1} \tanh(\tauhat/\phi)$.
Further integration yields
$$
\align
\zeta
 &= \ln (\mo \cosh (\tauhat/\phi))
\tag38 \\
\vspace{-7pt}
\intertext{and}
\vspace{-7pt}
\[[ x, y, z, t @,@,@,@, \]]
 &= \[[ x_0, y_0, z_0, t_0 \]]
       + \[[ a, b, c, E @,@,@,@, \]] \mo^{-2} \phi \tanh (\tauhat/\phi).
\tag39
\endalign
$$
Equations (32) are satisfied if $\tauo = \mo^{-1} \phi \tanh(\tauhat/\phi) =
q/\mo$.

The geodesic path $p$ that these equations describe is complete, in that
$p(\tauhat)$ exists for $-\infty < \tauhat < \infty$.  The test particle
experiences, therefore, a full, historically complete existence in
space-time-\!-time.  Contrarily, however, its sojourn in space-time is
constricted to the times between $t(-\infty)$ and $t(\infty)$, where
$t(\pm \infty) = t_0 \pm \L(E/\mo^2\R) \phi$.  In the eyes of a space-time observer
the particle, if its energy $E$ is positive, springs into full-blown existence
at the event $\eusm E_1$ whose coordinate vector is
$\[[ x, y, z, t \]](-\infty)$, traveling from the very instant of its birth
with uniform velocity toward the event $\eusm E_2$ with coordinate vector
$\[[ x, y, z, t \]](\infty)$, at which it vanishes, having lived a lifetime of
precisely the span given by $t(\infty) - t(-\infty) = 2 \L(\phi E / \mo^2\R)$
in coordinate time, and by $\tauo(\infty) - \tauo(-\infty) = 2 (\phi / \mo)$ in
its own proper time, a span that tends to $\infty$ as $\mo$ is decreased to 0.
This sudden appearance and disappearance is an artifact of the projecting of
the $\tauhat$-complete geodesic from the five dimensions of space-time-\!-time
onto the four of space-time.  It is entirely analogous to what happens when
geodesics are projected from the four dimensions of de Sitter's space-time onto
the three dimensions of space.  In the de Sitter case the projections terminate
at points of space, in the hyper-de Sitter case, at points of space-time, that
is, at events. 

The behavior of $q$ is particularly interesting.  Beginning with the asymptotic
value $-\phi$ at $\eusm E_1$, $q$ increases monotonically (linearly with
respect to $\tauo$, at the rate $\mo$), and finishes with the asymptotic value
$\phi$ at~$\eusm E_2$.  In the vacuum, where $F = 0$, this has no direct effect
on the particle motion, as the Lorentz force vanishes.  But in a nonvacuum,
where the charge evolution equation is still $\dot q = 1 - (q / \phi)^2$,
similar behavior, including sudden appearance of the particle at an event
$\eusm E_1$ and
disappearance at an event $\eusm E_2$, will persist generically, with the
notable consequence that the particle will respond to an electromagnetic field
initially as negatively charged with $q = -\phi(\eusm E_1)$, but ultimately as
positively charged with $q = \phi(\eusm E_2)$, passing through a state of
electrical neutrality at some intermediate event.  To explain, if one can, how
this behavior could be consistent with empirical observations will require a
detailed investigation not to be undertaken here.  Such an explanation clearly
would center on the specifics of the transition from the negatively charged
state to the positively charged state, particularly on how the ambient fields
affect the time and place of that transition.  That the undertaking can produce
remarkable dividends is suggested by the following considerations.

The coupling of the fourth force to the momentum rates in Eqs\. (33) and
($33'$) involves the factor~$e^{2 \zeta}$, not present in the couplings of the
first three.  At the terminal events $\eusm E_1$ and $\eusm E_2$ this factor
goes to $\infty$, with the singular result that the fourth force effectively
becomes infinitely strong, overwhelms the first three, and takes control of all
aspects of the particle's space-time trajectory except the precise timing of
its terminal events.  If the potential $\ln \phi$ has valleys, then at the two
ends of its space-time history the particle, no matter what it might do or
where it might go in the interim, will be forced down into one of those
valleys, to oscillate to and fro through its greatest depth with ever higher
frequency and ever smaller amplitude.  In more picturesque language, the
particle, though it wander hither and yon in midlife, must with high
probability be born shaking near a valley bottom and die trembling in a similar
place.  What is more, the electric charge of the particle at birth or at death
will be determined environmentally by the value of $\phi$ at the event in
question, thus will be more a characteristic of the space-time-\!-time than of
the individual particle.  That all this suggests the possibility of explaining
some aspects of atomic structure without invoking quantum theory is apparent.
By way of illustration it is interesting to contemplate the vacuum
space-time-\!-time modified so that $\phi = R e^{f(\rho)}$, where
$\rho = \sqrt {x^2 + y^2 + z^2}$.  The valleys of $\ln \phi$ bottom out at the
local minima of $f(\rho)$.  If, for example,
$f(\rho) = - \cos \L(2 \pi \sqrt {\rho / \rho_1}\,\R)$, the bottoms are
stationed where $\rho = n^2 \rho_1$, for $n = 0, 1, 2, 3, \ldots$, which for
$n =  1, 2, 3, \ldots$ mimics the spacing of circular electron orbits in the
Bohr model of the hydrogen atom if $\rho_1$ is the ground state radius.  A test
particle in this space-time-\!-time can live and die in one of these valleys
while another, born at the same place and time with the same electric charge
but with greater or lesser energy, migrates to some other valley to perform its
disappearing act.
\footnote{
A proper interpretation, based on the geometrical construction described in
{\cite 3}, is not that the test particle is ``born'' at $\eusm E_1$ and
``dies'' at $\eusm E_2$, but that it only {\it appears} at $\eusm E_1$ and
{\it disappears} at $\eusm E_2$.}

The vacuum equations of motion admit solutions with $\mo = 0$, $\epsilon = 1$,
$q = \pm \phi$, and $\dot p^\mu \neq 0$, thus admit massless charged test
particles traveling at the speed of light.  Such particles exist also in
nonvacuum space-time-\!-times of the same signature, but only in highly
restricted circumstances.  Each of them must follow a path confined to a level
surface of $\phi$, for if $\mo = 0$, then $\dot q = 0$, so $q$ must be both
constant and equal to $\pm \phi (p)$, and therefore
$\phi (p) = |q| = \text{const}$.  Moreover, satisfaction of the second of
Eqs\. (28) requires in general that if $\phi$ is time-independent, then
$d\phi$ must vanish on $p$ and the forces
corresponding to the first two terms inside the parentheses must balance one
another.  In the modified-vacuum example above, these requirements cannot be
met, as $d\phi = 0$ necessitates that the orbits be circular, and there is no
Lorentz force to balance the resulting centrifugal force term.  If, however,
the vacuum is further modified to include a Coulomb potential in the form
$\Ao = (Q/\rho) dt$ with $Q > 0$, then the requisite balance can be attained
with $q = - \phi (p)$.  In that space-time-\!-time will be found, therefore,
negatively charged, massless particles circulating at lightspeed in the valley
bottoms of $\phi$, where $\rho = n^2 \rho_1$, and (unstably) on the ridge
crests of $\phi$, where $\rho = \L(n + {1\over2}\R)^2 \rho_1$.  Also to be
found in those locations are charged, massless particles for which
$\dot p^\mu = 0$, their space-time existences confined to single events
$\eusm E$, their charges restricted to $q = \pm \phi (\eusm E)$, and their
$\zeta$ dependencies given by $\zeta = \zeta_0 + \tauhat/q$. Such test
particles can exist, in fact, in every space-time-\!-time at every space-time
event $\eusm E$, if any, at which $d\phi = 0$, but at no other event.

A $\phi @,@,@,$-wave form with a null propagation vector $v^\mu$ can carry
charged, massless particles from place to place at the speed of light with
${\dot p}^\mu \propto v^\mu$, provided as above that the $\Gammao$- and the
$F$-terms sum to zero in Eqs\. (28), as they do when, for example,
$\Gammao_\kappa {}^\mu {}_\lambda = 0$ and $F$ describes an electromagnetic
wave form with propagation vector $v^\mu$, so that
$F_{\kappa \lambda} {\dot p^\lambda} \propto F_{\kappa \lambda} v^\lambda = 0$.
A useful example to study is the vacuum modified so that $\phi = U(t - x)$ and
$\Ao = V(t - x) y (dt - dx)$, representing planar $\phi @,@,@,$-wave and
$\Ao$-wave forms propagating in the positive $x$ direction at lightspeed, with
$d\phi = U'(t - x) (dt - dx)$ and with
$F = 2 V(t - x) (dt \wedge dy - dx \wedge dy)$, a linearly polarized plane-wave
solution of the vacuum Maxwell equations.  With $\epsilon = \epshat = 1$,
$\mo = 0$, and $U$ nowhere constant, the equations of motion integrate to
$$
\align
\[[ \dot x, \dot y, \dot z, \dot t @,@,@,@, \]]
 &= \[[ 1, 0, 0, 1 @,@,@,@, \]] (a + b \tauhat) e^{-2 \zeta}
\tag40 \\
\vspace{-7pt}
\intertext{and}
\vspace{-7pt}
\zeta
 &= \zeta_0 + \tauhat/q,
\tag41
\endalign
$$
where $c = t - x$ (a constant of the motion), $a$ is a constant,
$b = U'(c)/U(c)$, and $q = \pm U(c)$.  When $b = 0$, $d\phi (p) = 0$, so the
particle rides along secure in a wave trough bottom of $\phi$ or balanced on a
wave crest or wave shoulder.  If $b > 0$, then $\dot t$ and $\dot x$ will
switch from positive to negative when $\tauhat = -a/b$, and $t(-a/b)$ and
$x(-a/b)$ will be maximum values.  Its five-dimen\-sional proper time $\tauhat$
increasing, the particle will cease to advance and begin to retreat in ordinary
(space-time) time, maintaining, however, the velocity
$\[[ dx/dt, dy/dt, dz/dt @,@,@,@, \]] = \[[ 1, 0, 0 @,@,@,@, \]]$.  A
space-time observer will perhaps interpret this as two particles on one track
that at a certain instant jointly vanish without a trace.  This variation on
the disappearing act takes place where $U'(t - x) > 0$, which puts the
particle(s) on the downward sloping front side of a $\phi$ wave.  If $b < 0$,
$t$ and $x$ will have minimum values, and the particle(s) will seem to appear
out of nowhere on the back side of a $\phi$ wave.

It is worth emphasizing that the electromagnetic field in this example exerts
no force on the charged, massless particles.  Because these ``charged photons''
are moving with the same velocity with which the electromagnetic field is
propagating, they are immune to its influence.  In the previous example, on
the other hand, the static Coulomb field supplies the only apparent force
(other than the even more fictitious centrifugal force) exerted on such
particles, but only because it holds them with precision in the valley bottoms
and on the ridge crests of $\ln \phi$, where the gradient force vanishes.

Let us note in passing that space-time-\!-time also provides geodesic paths for
massless, electrically neutral test particles traveling at the speed of light,
free of the bondage suffered by the particles discussed above.  They are the
paths on which $\epsilon = \mo = q = 0$.  According to Eq\. (28) these
particles are acted upon by the Einstein force, but not by any other of the
four ``forces'' identified in Sec\. 7.  The vector potential $\Ao$
does, however, affect them indirectly by way of the equation
$\dot \zeta = - \Ao_{\mu} \dot p^{\mu}$, which follows from Eqs\.~(26).  It is
only this subtle effect that can cause such a ``neutrino'' particle's
space-time history to have a beginning or an ending event, absent which the
particle would be unable to participate in an ``interaction'' of the kind I
shall now propose.

Several test particles of the types described above, including in particular
the charged, massless particles that disappear the instant they appear, can
have in common an event $\eusm E$ at which each either appears, with
$q = - \phi (\eusm E)$ or $q = 0$, or disappears, with $q = \phi (\eusm E)$ or
$q = 0$. They can be made to ``interact'' by demanding that the asymptotic
values of their kinematical variables obey a ``conservation law'' of some sort.
A natural candidate is this

\parindent=30pt
{\narrower\smallskip
\noindent {\bf Space-Time-\!-Time Conservation Law.}  The sum of the asymptotic
space-time-\!-time momenta at the space-time event $\eusm E$ of all the
particles whose space-time trajectories begin or end at $\eusm E$ is zero.
\smallskip}
\parindent=20pt

\noindent This law would comprise for such interactions both the conservation
of space-time four-momentum and the conservation of electric charge.
\eject

\vskip 12pt

\line{\bf 9.  REMARKS \hfil}
\vskip 3pt

\indent It is possible to look upon space-time-\!-time with its fields and test
particles as a purely geometric, deterministic substructure underlying quantum
theory, somewhat as the molecular structure of gases underlies their
thermodynamical theory.  The statistical indeterminacies and probabilistic
predictions characteristic of quantum theory would, on this view, arise from
the variability and preferential tendencies of asymptotic endings of individual
test particle tracks in space.  When two or more test particles in
thrall to a $\phi$ field interact at an event $\eusm E$, that event can occur
anywhere in space, but is more likely to happen near some one of $\ln \phi$'s
valley bottoms than off in the highlands.  That $\eusm E$ will occur exactly at
the bottom is, however, unlikely.  Even though an infinitely growing force
causes the particles to oscillate about the bottom depth with increasing
frequencies and diminishing amplitudes, only by the merest chance will they
terminate their space-time histories (thus consummate their interaction)
precisely there.  Instead, they will disappear from space-time at some nearby
point while the force is still trying to have its way with them --- the
magician's method of escape from bondage, so to speak.  A random selection of
such groups of interacting particles will produce a statistical cloud of
interaction events whose density will peak at the valley bottoms of $\ln \phi$.

The proposed space-time-\!-time conservation law speaks loosely of ``all the
particles whose space-time trajectories begin or end at $\eusm E$.''  In truth
no ``particles'' have been identified in space-time-\!-time, only geodesics to
be treated as possible paths of ``test particles.''  Of these geodesics
innumerably many have space-time projections that begin or end at $\eusm E$,
but only a few of those would be expected to have space-time projections that
terminate at a prescribed second event.
\footnote
{In the space-time-\!-time vacuum only one of them does, as the terminal events
$\eusm E_1$ and $\eusm E_2$ fully determine the integration constants in
Eqs\. (38) and (39).}
Someone setting out to analyze the ``interaction''
between a particle appearing or disappearing at one event and another particle
appearing or disappearing at another event, on the basis of particle
``exchanges'' (both direct and indirect via intermediate events), might soon
begin assembling such projections of geodesics into diagrams of the Feynman
type, using the space-time-\!-time conservation law as the guide to vertex
formation.

By making a test particle's electric charge $q$ at an interaction event
$\eusm E$ be determined solely by the ambient geometry, through the asymptotic
equation $q = \pm \phi (\eusm E)$, \text{space-time-\!-time} theory explains at one
stroke both the discreteness and the uniformity of electric charge, that is to
say, of the {\it passive} electric charge of particles that can be treated as
test particles.  The price of this is that every such particle for which
$\mo > 0$ must over the course of its lifetime watch its charge account balance
go inexorably from negative to positive.  On the face of it this would seem to
present a grave difficulty for any attempt to identify space-time-\!-time test
particles with, for example, electrons.  To deny that possibility at this stage
would, however, be premature.  The difficulty might be resolved if, for
instance, in regions of space-time where electron charges are actually
measured $\phi$ is essentially constant (equal to $e$ if $-e$ is the measured
electron charge), and an electron ejected from an atom and passing through such
a region maintains $q$ close to $-\phi$ while there but quickly runs its
charge balance up to $\phi$ as it is being captured by a target ---
provided, of course, that space-time-\!-time geodesics reflecting this behavior
can be found.  Whether such a resolution is realizable is a question demanding
further investigation.

The ambiguity involved in considering both $\epshat = 1$ and $\epshat = -1$ to
yield a legitimate space-time-\!-time metric $\Ghat$ can be resolved by
absorbing both cases into an enlarged geometry.  It suffices to expand $\zeta$
into a complex coordinate and $\phi$ and $\Ao$ into complex-valued fields.  To
keep the space-time metric $\Go$ real under coordinate gauge transformations
$\zeta \to \zeta - \lambda$, with $\lambda = \mu + i \nu$, requires
introduction of the additional transformation $\phi \to \phi e^{-i \nu}$.
Thus, if $\zeta' = \zeta - \lambda$, then $\Ghat = e^{i 2 \nu} \Ghat'$, where
$\Ghat' = e^{2 \zeta'} \Go' + {\phi'}^2 \L(\Ao' + d \zeta'\R) \otimes
\L(\Ao' + d \zeta'\R)$, with $\Go' = e^{2 \mu} \Go$, $\Ao' = \Ao + d \lambda$,
and $\phi' = \phi e^{-i \nu}$.  The phase shift in $\phi e^{-i \nu}$ is
reminiscent of London's electron wave field phase shift that Weyl embraced.$^4$

This enlargement of the geometry {\it via} partial complexification can be
accomplished for Kaluza--Weyl metrics in general by modifying the EEC to
specify that $\xi$ be a complex vector field (which entails that the
corresponding dimension of $\eusm M$ become complexified).  The EEC also lends
itself to modification in two less drastic ways:  alteration of the
condition $\eusm L_\xi \Ghat = 2 G$ to the less restrictive
(a) $\eusm L_\xi G = 2 G$ or the more restrictive
(b) $\eusm L_\xi \Ghat = 2 \Ghat$.

The effect of changing to (a) is to admit a $\zeta$-dependence of $\phi$ and of
$\Ao$.  This version of space-time-\!-time geometry is the one
described in \cite{3} (an extended description appears in \cite{12}).  What is
apparently the same or an equivalent geometry was studied from a projective
viewpoint by R. L. Ingraham, with results presented in a sequence of papers
that appeared mainly in {\it Il Nuovo Cimento}, beginning in 1952
(see \cite{13} and references therein).  The focus was primarily on the theory
of fields satisfying equations invariant under the conformal group of
Minkowskian space-time, these fields being defined on the five-dimensional
projective spaces whose points are the (hyper)spheres of Minkowskian
space-time.  To the limited extent that they can be compared, the physical
interpretations I have adopted and those of Ingraham differ appreciably.
Whereas Ingraham's address primarily the problem of deriving from the geometry
a concept of ``mass,'' the interpretations I have imposed on the
space-time-\!-time geometry speak to both the mass concept and the concept of
``electric charge,'' and call for a much more fundamental revision of the
latter concept than of the former.

Changing to (b) produces a metric $\Ghat$ of the form
$\Ghat = e^{2 \zeta} \bar G$, where
$\bar G = \Go + \epshat \phi^2 \L(\Ao + d\zeta\R) \otimes \L(\Ao + d\zeta\R)$
with $\Go$, $\phi$, and $\Ao$ independent of $\zeta$.  In this context the
coordinate change $\zeta \to \zeta - \lambda$ produces the usual metric gauge
transformation $\Go \to e^{2 \lambda} \Go$, but produces instead of the usual
gradient gauge transformation the mutilated version
$\Ao \to e^\lambda \L(\Ao + d\lambda\R)$, of no obvious utility.  A
five-dimensional metric like $\bar G$, but with $\phi^2 \Go$ in place of $\Go$,
and only $\Go$ and $\Ao$
independent of $\zeta$, was arrived at in a formalistic manner by
Vladimirov \cite{14}, who suggested for $\phi$ the forms
$\phi = \tilde \phi (x^\mu) \exp (i K \zeta)$ and $\phi = \exp (K \zeta)$,
with $K$ a real constant; these would correspond respectively to
$\eusm L_\xi \Ghat = 2 i K \Ghat$ and $\eusm L_\xi \Ghat = 2 K \Ghat$.  Unlike
the EEC modified by (a), which when $\phi$ and $\Ao$ are independent of $\zeta$
yields the same geometry as the unmodified EEC, thus can support the same
physical interpretations, the EEC modified by (b) pushes the Kaluza and the
Weyl geometries into an awkward and unnatural union, therefore seems
unlikely to become a source of deep insight into the foundations of physics.

\enddocument